\documentclass[aps,prl,twocolumn,showpacs,floatfix,nobibnotes,superscriptaddress,amsmath,amssymb,longbibliography]{revtex4-1}

\usepackage{graphicx}
\usepackage{epsfig}
\usepackage{bm}
\usepackage{color}
\usepackage{float}
\usepackage{dcolumn}
\usepackage{multirow} 
\usepackage[unicode=true,
  linktocpage,
  linkbordercolor={0.5 0.5 1},
  citebordercolor={0.5 1 0.5},
  colorlinks=true,
  linkcolor=blue,
  citecolor=blue,
  urlcolor=blue]{hyperref} 
\usepackage{cgp_macros}
\usepackage{lipsum} 
\usepackage{placeins} 

\graphicspath{{.}}

\newcommand{\beq}{\begin{equation}}
\newcommand{\eeq}{\end{equation}}
\newcommand{\beqa}{\begin{eqnarray}}
\newcommand{\eeqa}{\end{eqnarray}}

\newcommand{\bra}[1]{\left\langle #1 \right|}
\newcommand{\ket}[1]{\left| #1 \right\rangle}
\newcommand{\braket}[2]{\left\langle #1 \vert #2 \right\rangle}

\newcommand{\NNLOsat}{NNLO$_{\rm sat}$}
\newcommand{\NNLOgod}{$\Delta$NNLO$_{\rm GO}$(450)}

\newcommand{\CEnuNS}{CE$\nu$NS}
\newcommand{\Rp}{R_\text{p}}
\newcommand{\Rn}{R_\text{n}}
\newcommand{\Rskin}{R_\text{skin}} 
\newcommand{\chiEFT}{$\chi$-EFT}

\begin{document}

\title{Coherent elastic neutrino-nucleus scattering on \tAr{40} from first principles}

\author{C.~G.~Payne}
\affiliation{Institut f\"ur Kernphysik and PRISMA$^+$ Cluster of Excellence, Johannes Gutenberg-Universit\"at, 55128
  Mainz, Germany}

\author{S.~Bacca}
\affiliation{Institut f\"ur Kernphysik and PRISMA$^+$ Cluster of Excellence, Johannes Gutenberg-Universit\"at, 55128
  Mainz, Germany}

\author{G.~Hagen}
\thanks{This manuscript has been authored by UT-Battelle, LLC under
  Contract No. DE-AC05-00OR22725 with the U.S. Department of
  Energy. The United States Government retains and the publisher, by
  accepting the article for publication, acknowledges that the United
  States Government retains a non-exclusive, paid-up, irrevocable,
  world-wide license to publish or reproduce the published form of
  this manuscript, or allow others to do so, for United States
  Government purposes. The Department of Energy will provide public
  access to these results of federally sponsored research in
  accordance with the DOE Public Access
  Plan. (http://energy.gov/downloads/doe-public-access-plan).}
\affiliation{Physics Division, Oak Ridge National Laboratory,
Oak Ridge, TN 37831, USA} 
\affiliation{Department of Physics and Astronomy, University of Tennessee,
Knoxville, TN 37996, USA} 

\author{W.~Jiang} 
\affiliation{Department of Physics and Astronomy, University of Tennessee,
Knoxville, TN 37996, USA} 
\affiliation{Physics Division, Oak Ridge National Laboratory,
Oak Ridge, TN 37831, USA} 

\author{T.~Papenbrock}
\affiliation{Department of Physics and Astronomy, University of Tennessee,
Knoxville, TN 37996, USA} 
\affiliation{Physics Division, Oak Ridge National Laboratory,
Oak Ridge, TN 37831, USA}

\begin{abstract}
Coherent elastic neutrino scattering on the \tAr{40} nucleus is
computed with coupled-cluster theory based on nuclear Hamiltonians
inspired by effective field theories of quantum chromodynamics. Our approach 
is validated by calculating the charge form factor and comparing it to data from electron
scattering. We make predictions for the weak form factor, the neutron radius, 
and the neutron skin, and estimate systematic uncertainties. The neutron-skin
thickness of \tAr{40} is consistent with results from density functional
theory. Precision measurements from coherent elastic neutrino-nucleus scattering could potentially be used 
to extract these observables and help to constrain nuclear models.
\end{abstract}

\maketitle 

{\it Introduction.}--Fundamental properties of atomic nuclei, such as
the distribution of the protons within the nucleus, are well
determined from electron scattering experiments. In contrast, the
distribution of the neutrons within the atomic nucleus, an equally
important and fundamental property, is not as well known because it is
difficult to measure. Parity violating electron scattering
experiments~\cite{PREX,PREX_CREX,Mainz_Review} offer the least model dependent
approach to experimentally probing the neutron
distribution. Other processes occurring through neutral current weak
interactions, i.e., by the exchange of a $Z^0$ boson, may offer an alternative
and attractive opportunity in the future. A prominent example is
coherent elastic neutrino-nucleus scattering (\CEnuNS), a process which is 
sensitive to the neutron distribution and the neutron
radius~\cite{Amanik_2008,Patton2012,Cadeddu}. Even though neutrinos
are notoriously elusive particles, the COHERENT collaboration recently
observed, for the first time, 
\CEnuNS\ from a sodium-doped CsI
detector~\cite{Akimov1123}. The experiment used stopped-pion
neutrinos~\cite{Scholberg2006} from the Spallation Neutron Source at
Oak Ridge National Laboratory, and discovered \CEnuNS\ at a
$6.7\sigma$ confidence level with neutrinos coming from: delayed
electron neutrinos, muon anti-neutrinos, and prompt
muon neutrinos. The next stage of the COHERENT experiment is to switch
to a $\orderof\!1$ ton target of liquid argon. Liquid argon will also
be used in the future long-baseline neutrino experiment
DUNE~\cite{DUNE}, which is aimed at extracting neutrino parameters from the
observation of their oscillations. In addition, liquid argon is being used for a number of dark matter experiments (DEAP-3600~\cite{deap}, DarkSide~\cite{Darkside}, ArDM~\cite{ArDM}, MiniCLEAN~\cite{miniclean}), for which coherent neutrino scattering is important to determine the so-called neutrino floor.
Studying the properties of the
\tAr{40} nucleus, the most abundant argon isotope composing the above
mentioned detectors, is thus an important task for
nuclear theory.

In the past decade we have seen an impressive progress in the theoretical
and computational tools that underpin our understanding of the nucleus
as a compound object of interacting protons and neutrons. A number of
ab initio calculations of nuclear electroweak properties that start
from interactions and currents obtained from chiral effective field
theory have successfully described key observables, see, e.g.,~\cite{BaccaPastore2014,lovato2014,pastore2017,gysbers2019,Barbieri:2019ual}.
The level of accuracy and confidence reached by ab initio calculations
in light- and medium-mass nuclei, along with the ability to access increasing mass
numbers, allows us to address
open questions in neutrino physics. This makes a first principles
investigation of the \tAr{40} nucleus both urgent and timely. For instance, neutrino elastic scattering has been discussed as a way to
access the neutron-skin thickness~\cite{Cadeddu}, thus making it interesting
to compute this quantity in \tAr{40}. The neutron-skin thickness
impacts the equation of state of infinite-nuclear matter and has
astrophysical implications~\cite{Lattimer2012}.

In this Letter we compute the nuclear
weak form factor and the neutron-skin thickness with coupled-cluster theory from first principles,
and provide theoretical predictions that may eventually be probed experimentally.

{\it Coherent scattering.}-- Coherent elastic neutrino-nucleus scattering occurs in the regime $qR \ll 1$. Here $q=|{\bf q}|$ is the
absolute value of the three-momentum transfer from the neutrino to the
nucleus, and $R$ is the weak nuclear radius. In this regime, the neutrino
scatters coherently from the constituents of the nucleus, i.e., $Z$
protons and $N$ neutrons. The \CEnuNS\ cross section is
\begin{eqnarray}
\label{eq_cs}
\frac{d\sigma}{dT}(E_\nu,T) &\simeq& \frac{G_F^2}{4\pi} M \left[ 1- \frac{MT}{2E_\nu^2} \right] Q^2_W F^2_W(q^2) \,.
\end{eqnarray}
Here $G_F$ is the Fermi constant, $M$ is the mass of the nucleus, $E_\nu$ is the energy of the neutrino beam, and $T$ is the nuclear recoil energy. The weak charge $Q_W$ and weak form factor $F_W(q^2)$ are defined as
\begin{eqnarray}
\label{eq_w}
Q_W &=& N -(1-4\sin^2\theta_W)Z \,,  \\
\nonumber
 F_W(q^2) &=& \frac{1}{Q_W}\left[ NF_n(q^2) -(1-4\sin^2\theta_W)ZF_p(q^2) \right] \,,
\end{eqnarray}
 respectively. Here $\theta_W$ is the Weinberg weak mixing angle, and
 $F_{n,p}(q^2)$ is the proton ($p$) and neutron ($n$) form factor,
 respectively. Using the low-energy value of
 $\theta_W$~\cite{PDG2018} from the Particle Data Group, one obtains
 $1-4\sin^2\theta_W(0) = 0.0457 \pm 0.0002$. Thus, the weak
 form factor becomes $F_W(q^2)\simeq F_n(q^2)$, and
 \CEnuNS\ is mainly sensitive to the distribution of neutrons within the
 nucleus. The resulting cross section scales as $N^2$. In this paper
 we will consider low-$q$ ranges and investigate effects due to the
 nuclear structure. For \tAr{40}, the coherence condition limits $q
 \lesssim 50$ MeV/c, but we are also interested in exploring the form
 factors as ground-state observables in a wider momentum range. 

{\it Method.}-- Our computations are based on coupled cluster 
theory~\cite{coester1958,coester1960,kuemmel1978,bishop1991,mihaila2000b,dean2004,kowalski2004,bartlett2007,hagen2014},
where one solves the Schr\"{o}dinger equation
\begin{equation}
\overline{H}_N \vert \Phi_0\rangle = E \vert \Phi_0\rangle
\end{equation}
based on the reference state $\vert \Phi_0\rangle$ of a closed-shell nucleus. The similarity
transformed Hamiltonian is
\begin{equation} 
\overline{H}_N= e^{-T} H_N e^T\,.
\end{equation} 
The Hamiltonian $\overline{H}_N$ is normal-ordered with respect to the
reference state. The operator $T = T_1 + T_2 + T_3 + \ldots$ is
expanded in particle-hole excitations with respect to the reference
and is truncated at some low-rank particle-hole excitation level.
Following Ref.~\cite{Mirko}, we will denote coupled-cluster singles
and doubles calculations (where $T = T_1 + T_2 $) with ``D'', while
calculations that include linearized triples will be labeled with
``T-1''; we refer the reader to that paper and the
review~\cite{hagen2014} for details on the accuracy of various
coupled-cluster approximations in nuclei.

The open-shell nucleus \tAr{40} has
$Z=18$ protons and $N=22$ neutrons. We calculate its ground state using a
double-charge-exchange equation-of-motion technique~\cite{liu2019} starting from the closed-shell nucleus
$^{40}$Ca. This technique is a generalization of
single-charge exchange, used previously to describe the
daughter nuclei resulting from $\beta$-decays of closed-shell nuclei
such as $^{14}$C~\cite{ekstrom2014} and $^{100}$Sn~\cite{gysbers2019}.
The double-charge-exchange operator
\begin{eqnarray}
  R &=& {1\over 4}\sum_{p,p',n,n'} r_{pp'}^{nn'}\hat{n}^\dagger \hat{n'}^\dagger \hat{p'}\hat{p} \nonumber\\
  &&+ {1\over 36}\sum_{N,N',p,p',n,n'} r_{N'pp'}^{Nnn'}\hat{n}^\dagger \hat{n'}^\dagger \hat{N}^\dagger \hat{N'}\hat{p'}\hat{p} 
\end{eqnarray}
generates the ground-state of \tAr{40} as an excitation of the
$^{40}$Ca. Here, $\hat{p}$, $\hat{n}$, and $\hat{N}$ annihilate a
proton, neutron, and nucleon, respectively. The excitation amplitudes
$r_{pp'}^{nn'}$ and $r_{N'pp'}^{Nnn'}$ are solutions of the eigenvalue
problem $\overline{H}_N R = ER$, and the lowest eigenvalue $E$ is the
ground-state energy of \tAr{40}. Likewise, we define a left
excitation operator $L$ and also solve $L \overline{H}_N = EL$
(because $\overline{H}_N$ is not Hermitian). This allows us to evaluate
ground-state expectation values (such as the density) of operators
$\hat{O}$ as $\langle L |\overline{O}|R\rangle$. Here, $\overline{O}$ 
is the similarity transform of the operator $\hat{O}$. 

The computations shown in this
work are based on a model space that includes 15 major shells (unless otherwise specified) and an harmonic oscillator parameter $\hbar\Omega=16$ MeV. When we include leading triples T-1, we use an energy truncation $E_{\rm 3max}$ cut at 18 oscillator spacings, where we reach a sub-percentage convergence of the form factors in the considered momentum range.
 
\begin{figure}[hbt]
  \includegraphics[width=0.8\textwidth]{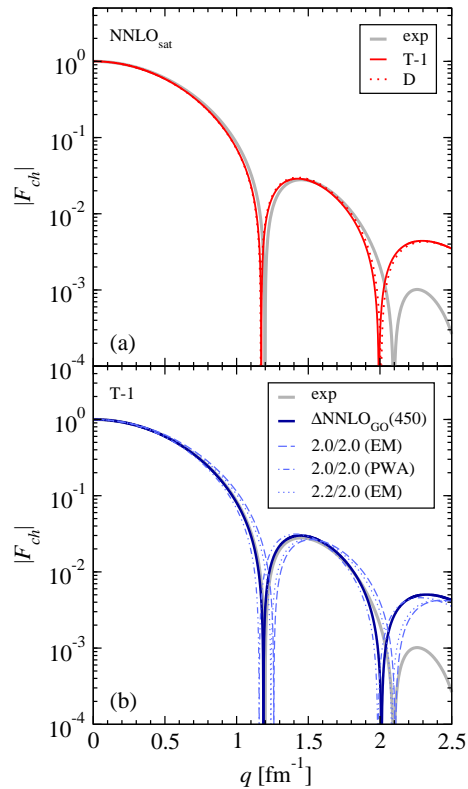}
  \caption{Panel (a): \tAr{40} charge form
    factor computed with the \NNLOsat\ interaction at two different levels
    of correlations (D and T-1), compared to experimental data (exp) by Ottermann {\it
      et al.}~\cite{Ottermann}. Panel (b): \tAr{40} charge form factor
    computed with various different interactions at the T-1 level, also compared to the experimental data. See text for more details.}
  \label{Fch}
\end{figure}

{\it Interactions.}-- We employ Hamiltonians from 
chiral effective field theories (\chiEFT) of quantum
chromodynamics (QCD)~\cite{vankolck1999,epelbaum2009,machleidt2011}.
In this framework, Hamiltonians are expressed in terms of nucleons and pions
and are consistent with the symmetries and broken chiral symmetry of QCD.
They are expanded in powers of $(Q/\Lambda_\chi)^{\nu}$, where $Q$ is
the low-momentum scale characterizing nuclear physics, and
$\Lambda_\chi\sim 1$ GeV is the QCD scale. The coefficients of the
Hamiltonian expansion are low-energy constants (LECs); they encapsulate the
unresolved short-range physics and are typically calibrated by adjusting 
theoretical results to experimental data. The accuracy of a calculation is controlled by the
order $\nu$ of the employed dynamical ingredients and by the accuracy
to which one can solve the many-body problem. In this work we
implement Hamiltonians derived at next-to-next-to-leading order or
higher ($\nu=3$ or $4$). To probe the systematic uncertainties, we
employ various chiral potentials. In particular, we use the \NNLOsat~ interaction~\cite{ekstrom2015},
for which the LECs entering the two-body and three-body forces are
adjusted to nucleon-nucleon phase shifts and to
energies and charge radii of light nuclei. We also use the \NNLOgod\ potential~\cite{jiang2019}, 
a delta-full \chiEFT~ interaction at next-to-next-to-leading order~\cite{ekstrom2018}, which was adjusted to light nuclei, and the saturation point and symmetry energy of nuclear matter. 
Finally, we employ selected soft potentials obtained
by performing a similarity renormalization group
transformation~\cite{bogner2007} of the two-body chiral potential by Entem and Machleidt~\cite{entem2003}, with leading-order three-nucleon forces
from \chiEFT~ adjusted to the binding energy of $^3$H and the charge
radius of $^4$He~\cite{nogga2004,Hebeler2011}. For these interactions we follow the notation of Ref.~\cite{Hebeler2011}, namely 1.8/2.0, 2.0/2.0, 2.2/2.0 (EM) and 2.0/2.0 (PWA), where the first (second) number indicates the cutoff of the two-body (three-body) force in fm$^{-1}$, and EM indicates that the pion-nucleon LECs entering the three-nucleon force are taken from the Entem and Machleidt potential~\cite{entem2003}, while in PWA they are taken from partial wave analysis data.
For electroweak operators we take the one-body terms, as two-body currents are expected to be negligible~\cite{lovato2013,Piarulli2013}, especially so at the low momenta of \CEnuNS.

\begin{figure}[hbt]
 \includegraphics[width=0.8\textwidth]{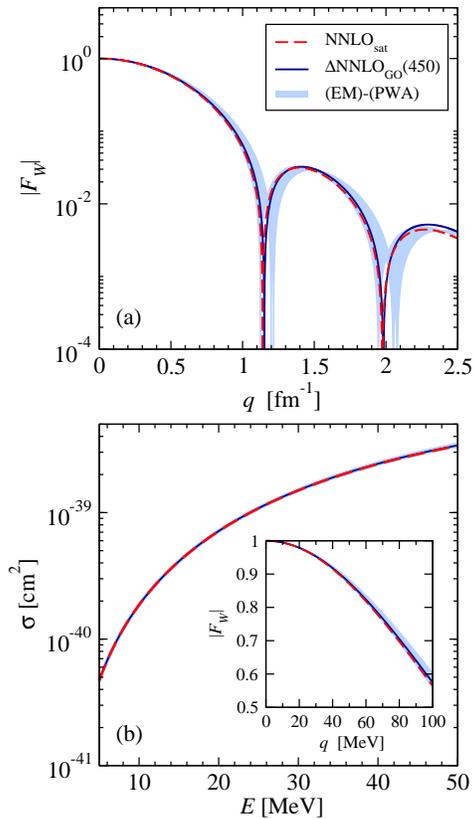}
  \caption{ Panel (a): \tAr{40} weak form factor
    computed with different Hamiltonians. The EM-family interactions are shown as a band. Panel
    (b): \CEnuNS\ as a function of the neutrino energy, computed with
    same three different Hamiltonians. The inset shows the form
factor zoomed into the low-$q$ region relevant to coherent scattering, in linear scale.
}
  \label{Fw}
\end{figure}

\begin{figure*}[hbt]
\includegraphics[width=0.99\textwidth]{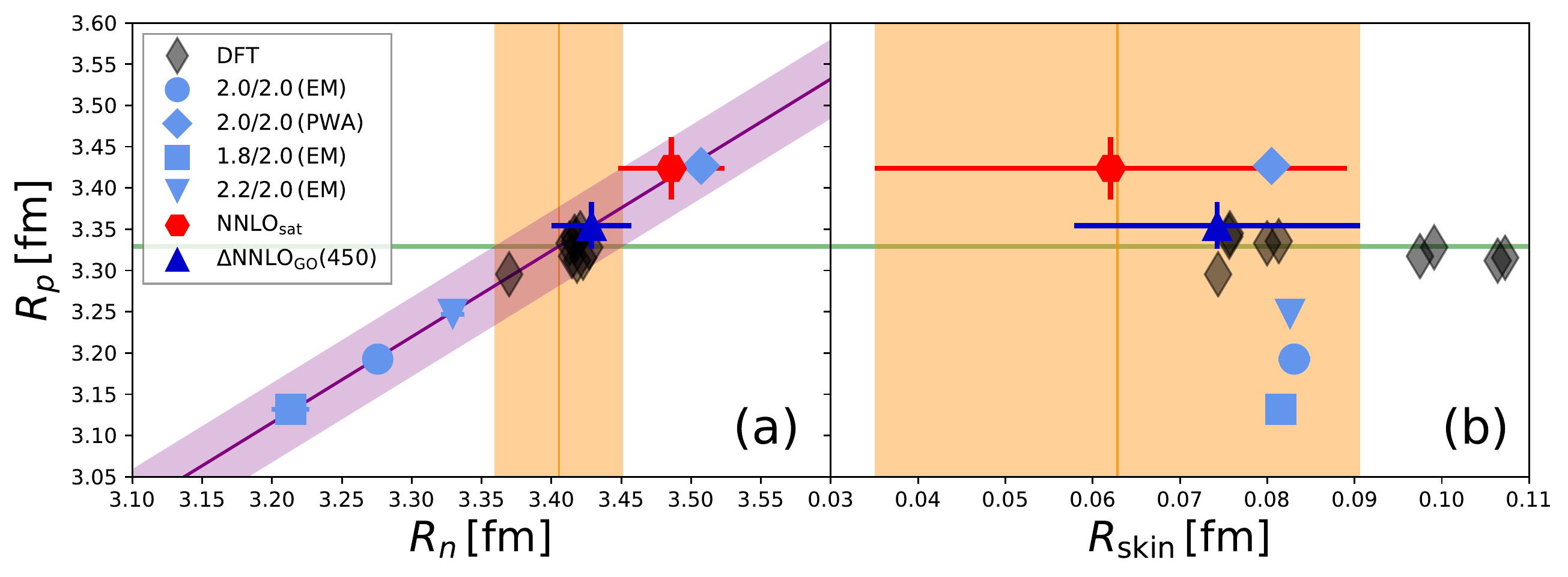}
  \caption{Correlation between $R_p$ and $R_n$ (a) and between $R_p$ and $R_{\rm skin}$ (b) for various Hamiltonians. The experimental $R_p$ is also shown by the horizontal green line~\cite{angeli2013}, as well as the DFT data~\cite{Schunk} by the diamonds.}
  \label{radii}
\end{figure*}

{\it Results.} -- Figure~\ref{Fch} shows our results for
the \tAr{40} charge form factor $F_{ch}$ as a function of $q$, and
compares them to electron-scattering data from Ottermann {\it et al.}~\cite{Ottermann}. This comparison validates
the theory. Panel (a) shows results from the \NNLOsat\ interaction for different
correlation levels of the coupled-cluster expansion. We see that increasing the
correlations from D to T-1 changes the form factor only slightly, and the results are sufficiently well converged. This is consistent with results from previous studies~\cite{Mirko,Simonis}, where triples correlations
only affected the radii below 1$\%$.
Panel (b) 
shows calculations of the charge form factor at the T-1 level for
different interactions. As representative examples we chose the 2.0/2.0 (EM), 2.0/2.0 (PWA), and 2.2/2.0 (EM) potentials.
 The form factors exhibit a dependence on the choice of the Hamiltonian, particularly at larger momentum transfers. 
The interaction \NNLOgod, derived in a delta-full chiral framework, provides a qualitatively similar description of the experimental data as the \NNLOsat, noting that the former interaction reproduces the first minimum of $|F_{ch}|$ more precisely.
We remind the reader 
that -- within the Helm model~\cite{helm1956} -- the 
first zero of the form factor is proportional to the 
inverse radius of the charge distribution.
Among the family of EM potentials, the 2.2/2.0 (EM) interactions predicts the first zero at
higher $q$, consistent with a smaller charge radius. Overall, one should trust the Hamiltonians only
for momentum transfers up to about $q=2.0$ fm$^{-1}$, which marks the scale of the employed ultraviolet cutoffs.

Figure~\ref{Fw}(a) shows the \tAr{40} weak form factor $F_W$ of
Eq.~(\ref{eq_w}) as a function of the momentum transfer $q$, calculated in the T-1 scheme. Here, we show the soft interactions with a band that encompasses the three different potentials, labeled with (EM)-(PWA).
The weak form factor exhibits a mild dependence on the choice of the Hamiltonian. The band spanned
by the from factors of the EM interactions exhibits a first dip 
at a larger $q$ value than 
the potentials \NNLOsat\ and the \NNLOgod, which are very similar. 
Our
results are consistent with a Helm form factor parameterized by a box radius of 3.83~fm and a surface thickness of 0.9~fm~\cite{Cadeddu}.
We also note that our ab initio results for the weak form factor agree 
with calculations from density functional theory~\cite{Patton2012}.

Let us consider the \CEnuNS\ cross section. Figure~\ref{Fw}(b) shows the
cross section calculated from Eq.~(\ref{eq_cs}) via $q^2 = \sqrt{2E_{\nu}MT/(E_{\nu} - T)} \approx \sqrt{2MT}$, as a function of the
neutrino beam energy, for three different interactions. The results are virtually independent of the employed potential, because only the
low-momentum part of the weak form factor contributes to the cross section.
The inset of Fig.~\ref{Fw}(b) shows the weak form factor for momentum transfers relevant to the coherent elastic neutrino-nucleus scattering. Even on the shown linear scale, one observes only a mild nuclear-structure dependence. For example, at $q=50$ and 100~MeV, $F_W$ has a $2\%$ and 6\% spread, respectively.
Consequently, \CEnuNS\ is required to reach a high precision in order to probe
differences in nuclear Hamiltonians. We remind the reader that the \CEnuNS\ signal scales with $N^2$, possibly making heavier nuclei such as caesium or iodine more attractive detector materials for this purpose than \tAr{40}.

Overall, the weak form factor has a very similar shape to the charge form factor. For the NNLO$_{\rm sat}$ interaction, at $q=0.25$ fm$^{-1}$ (1 fm$^{-1}$) $F_W$ is 
0.5$\%$ (20$\%$) smaller than $F_{ch}$, while the first dip of $F_W$ falls about 0.035 fm$^{-1}$ earlier than that of $F_{ch}$, meaning the neutron distribution extends further out from that of the protons.

We now turn to the computation of the point-proton $\Rp$ and point-neutron $\Rn$ radii for
\tAr{40}, as well as its neutron-skin thickness, defined as $\Rskin=R_n-R_p$. 
Figure~\ref{radii} shows the results obtained with T-1 coupled-cluster calculations for six different potentials. We employ the five previous ones and one other
member of the EM-interaction family~\cite{Hebeler2011}, namely the 1.8/2.0~(EM) interaction. 
The uncertainties of $R_p$ and $R_n$ are the difference between a T-1 and a D coupled-cluster theory calculation, and we take the maximum of the two values as the uncertainty for both. 
Our model space consists of 15 oscillator shells, except for the softest 1.8/2.0~(EM), which was already converged in 11 shells. As expected, uncertainties are larger for the harder interactions \NNLOsat~ and \NNLOgod.

As previously reported for $^{48}$Ca~\cite{hagen2015}, Fig.~\ref{radii}(a) also shows 
a strong correlation between $\Rp$ and $\Rn$. The
spread of the radii due to the variation of the employed Hamiltonians is about
$10\%$. As in Ref.~\cite{hagen2015}, a narrower constraint can be provided by intersecting the
correlation band -- obtained by linearly joining all our calculations with a symmetric spread (in purple) given by the maximum uncertainty bar -- with the
experimental value on $R_p$ taken from~\cite{angeli2013}. This yields $3.36\le R_n \le 3.45$~fm. Results from density functional theory~\cite{Patton2012,Schunk} are shown as the diamonds in Fig.~\ref{radii}(a).
These are all clustered around our constraint for $R_n$. Within uncertainties, our charge radius is also consistent with the recent ab initio computations of Ref.~\cite{Barbieri:2019ual}.

Results for the neutron skin are shown in Fig.~\ref{radii}(b). Because the neutron and proton radii are strongly correlated, the variation in $R_{\rm skin}$ is much reduced. The uncertainty of $R_{\rm skin}$ is the difference between the T-1 and D coupled-cluster computations. We predict the neutron-skin thickness of \tAr{40} in the range $0.035\!-\!0.09\ufm$. The results from density functional theory~\cite{Patton2012,Schunk} are again shown as diamonds. 
While consistent with the ab initio computation, we see that density functional theory predicts a slightly larger neutron-skin thickness. 

{\it Summary.}-- We performed calculations of the \tAr{40} charge and
weak form factors and observed a dependence on the choice of the
employed Hamiltonian, which is mild at low-$q$ and moderate in the
region of the first diffraction minimum. 
From the weak form factor, we calculated the
coherent elastic neutrino-nucleus scattering and observed that the Hamiltonian
dependence is probably too small to be disentangled by the COHERENT
experiment. On the other hand, we also provide predictions for the
neutron-, proton-, and neutron-skin thickness by exploiting the
correlations of coupled-cluster computations with various Hamiltonians
with the experimental value of $R_p$. The computed $R_n$ and $R_{\rm
 skin}$ of \tAr{40} are consistent with results from density functional theory,
and \CEnuNS\ with much improved precision could help to constrain
Hamiltonians from \chiEFT.

\begin{acknowledgments}
This work was supported 
 by the Deutsche
Forschungsgemeinschaft (DFG) through the Collaborative Research Center
[The Low-Energy Frontier of the Standard Model (SFB 1044)], and
through the Cluster of Excellence ``Precision Physics, Fundamental
Interactions, and Structure of Matter" (PRISMA$^+$ EXC 2118/1) funded by the
DFG within the German Excellence Strategy (Project ID 39083149), by the
 Office of Nuclear Physics,
U.S. Department of Energy, under grants desc0018223 (NUCLEI SciDAC-4
collaboration) and by the Field Work Proposal ERKBP72 at Oak Ridge
National Laboratory (ORNL).
Computer time was provided by the
Innovative and Novel Computational Impact on Theory and Experiment
(INCITE) program. This research used resources of the Oak Ridge
Leadership Computing Facility located at ORNL, which is supported by
the Office of Science of the Department of Energy under Contract
No. DE-AC05-00OR22725.

\end{acknowledgments}

\FloatBarrier 
\bibliography{master,refs,refs_SB} 

\end{document}